\documentstyle[twoside,fleqn,espcrc2]{article}

\newcommand{\AmS}{{\protect\the\textfont2
  A\kern-.1667em\lower.5ex\hbox{M}\kern-.125emS}}
\newcommand{\thin}{\hspace{0.04cm}}

\begin{document}

\pagestyle{empty}

\begin{flushleft}
\large
{SAGA-HE-149-99
\hfill May 31, 1999}  \\
\end{flushleft}
 
\vspace{2.3cm}
 
\begin{center}
 
\LARGE{{\bf Structure functions in polarized}} \\
\vspace{0.2cm}

\LARGE{{\bf proton-deuteron Drell-Yan processes}} \\

\vspace{1.8cm}
 
\LARGE
{S. Kumano $^*$} \\
 
\vspace{0.4cm}
  
\LARGE
{Department of Physics}         \\
 
\LARGE
{Saga University}      \\
 
\LARGE
{Saga 840-8502, Japan} \\

\vspace{2.0cm}
 
\Large
{Talk at the 7th international workshop} \\

\vspace{0.2cm}

{on Deep Inelastic Scattering and QCD} \\

\vspace{0.5cm}

{Zeuthen, Germany, April 19 -- 23, 1999 } \\

\vspace{0.05cm}

{(talk on April 22, 1999)}  \\
 
\end{center}
 
\vspace{1.8cm}

\vfill
 
\noindent
{\rule{6.0cm}{0.1mm}} \\
 
\vspace{-0.3cm}
\normalsize
\noindent
{* Email: kumanos@cc.saga-u.ac.jp.
     Information on his research is available at} \\

\vspace{-0.44cm}
\noindent
{\ \ \ 
http://www2.cc.saga-u.ac.jp/saga-u/riko/physics/quantum1/structure.html.} \\

\vspace{+0.5cm}
\hfill
{\large to be published in Nuclear Physics B}

\vfill\eject
\setcounter{page}{1}
\pagestyle{plain}


\title{Structure functions in polarized
           proton-deuteron Drell-Yan processes}

\author{S. Kumano
\address{Department of Physics, Saga University, 
                 Saga 840-8502, Japan} 
\thanks
{http://www2.cc.saga-u.ac.jp/saga-u/riko/physics
/quantum1/structure.html.
 S.K. was partly supported by the Grant-in-Aid for Scientific Research
from the Japanese Ministry of Education, Science, and Culture under
the contract number 10640277.}}


\begin{abstract}
Polarized proton-deuteron (pd) Drell-Yan processes are investigated
for studying new spin structure of a spin-1 hadron.
Our formalism indicates that there exist many structure
functions: 108 in general and 22 in the $\vec Q_T$-integrated case.
A naive parton model suggests that at least one of them should be related
to the tensor polarized distributions, which have not been measured
at all. There are some experimental possibilities at FNAL, HERA,
and RHIC to study the polarized pd reactions.
\end{abstract}

\maketitle

\section{Introduction}

Spin structure of the spin-1/2 nucleon has been investigated
extensively. On the other hand, spin structure of spin-1 hadrons
has not been well investigated in connection with new spin physics,
namely the tensor structure.
We know that the tensor structure function $b_1$ exists 
for a spin-1 hadron and it will be measured in the polarized
electron-deuteron scattering. However, it is known in the unpolarized
reactions that we can not determine the antiquark distributions
in the medium-$x$ region by the electron scattering data.
They are determined by the Drell-Yan measurements.
In the same way, the studies of polarized proton-deuteron (pd)
Drell-Yan processes should be valuable for finding the tensor-polarized
antiquark distributions. We discuss a general formalism
\cite{hk1} and a parton-model analysis \cite{hk2}
of the pd Drell-Yan processes. In particular, we explain
how a quadrupole spin asymmetry is related to the $b_1$-type
distributions.

\section{Possible structure functions}

The general formalism of the proton-proton (pp) Drell-Yan process
was completed many years ago, and it is the foundation
of the RHIC-SPIN project. On the other hand, the polarized pd formalism
was studied only recently \cite{hk1}. 
Because the space is limited,
only the major points are explained in this section.
Two independent methods are employed for finding possible structure
functions.

In the first method, the Jacob-Wick helicity formalism is used
by introducing the spin density matrix. The important point in
the formalism is that there exist rank-two tensors in the 
density matrix because of the spin-1 nature of the deuteron.
Imposing Hermiticity, parity conservation, and time-reversal
invariance, we find that 108 structure
functions exist in the pd Drell-Yan processes.
In comparison with the 48 functions in the pp Drell-Yan, there are 60
new structure functions. Of course, all of them are associated
with the rank-two terms, namely the tensor structure of the deuteron.
The 108 functions are too many to be investigated seriously.
In order to extract the essential ones, the cross section is
integrated over the virtual-photon transverse momentum $\vec Q_T$.
Even in this case, we find that there are 22 structure functions.
Because only 11 functions exist in the pp reactions, there are
11 additional ones. Therefore, the interesting point of
studying the pd Drell-Yan is to investigate these new structure
functions.

In the second method, the hadron tensor is expanded in terms
of possible combinations of momentum and spin vectors with
the conditions of Hermiticity, parity conservation, 
time-reversal invariance, and current conservation.
It is not useful to list all the 108 combinations; therefore,
only the $Q_T\rightarrow 0$ limit is considered in this formalism.
We have to be particularly careful about including spin-dependent
tensor terms in the expansion.
Assigning a structure function for each expansion coefficient,
we find also the 22 structure functions. It means that the possible
functions are confirmed by the two independent methods.
The details of these formalisms are found in Ref. \cite{hk1}.

The new structure functions are characterized by the polarizations
given by the spherical harmonics $Y_{20}$, $Y_{21}$, and $Y_{22}$. 
We express these quadrupole polarizations as $Q_0$, $Q_1$, and $Q_2$:
\begin{eqnarray}
Q_0  \! \! \! \! \! \! 
      & & \textstyle{\rm for\ the\ term\ }
                    \ 3 \, cos^2 \beta -1 \sim Y_{20} 
\ ,
\nonumber \\
Q_1  \! \! \! \! \! \! 
      & & \textstyle{\rm for} \hspace{1.7cm}  
                  sin \beta \, cos \beta \sim Y_{21} 
\ , \\
Q_2  \! \! \! \! \! \! 
      & & \textstyle{\rm for} \hspace{1.7cm} 
                  sin^2 \beta \sim Y_{22} 
\ ,
\nonumber
\end{eqnarray}
where $\beta$ is the polar angle of the spin polarization.
They are related to the quadrupole spin asymmetries in the $xz$, $yz$,
and $xy$ planes, respectively.
A $Q_0$ type structure function is measured by the difference
between the longitudinally and transversely-polarized cross sections.
A $Q_2$ one is measured by the difference between
the cross sections with the polarizations of $x$ and $y$ directions.
The $Q_1$ type structure functions are interesting in the sense
that they cannot be measured in the longitudinally-polarized
($\beta=0$) and transversely-polarized ($\beta=\pi/2$) reactions.
The optimum way of measuring them is to choose the polarization
angle in between ($\beta=\pi/4$).
In this sense, it may be called ``intermediate" polarization.
It is an important finding of our studies that there exist
intermediate structure functions which are not related to the
longitudinal and transverse polarizations.

The polarized structure functions could be obtained by polarization
asymmetry measurements. It is well known that five spin combinations
should exist in the pp Drell-Yan: unpolarized cross section
$< \! \sigma \! >$, longitudinal (transverse) double spin asymmetry
$A_{LL}$ ($A_{TT}$), longitudinal-transverse spin asymmetry $A_{LT}$,
and transverse single spin asymmetry $A_T$ (or denoted as $A_N$).
We should be careful in defining the spin asymmetries in the deuteron
reactions so that the tensor distributions should be excluded
from the denominator \cite{hk1}.
In the pd Drell-Yan, there are additional quadrupole asymmetries
and we have the following fifteen spin combinations
\begin{eqnarray}
& & \! \! \! \! \! \! \! \! \! \! \! \! \! \! \! \! \! 
< \! \sigma \! >, \thin A_{LL},   \thin A_{TT},   \thin A_{LT}, 
                  \thin A_{TL},   \thin A_{UT},   \thin A_{TU},
                  \thin A_{UQ_0},
\nonumber \\
& & \! \! \! \! \! \! \! \! \! \! \! \! \! \! \! \! \!    
                 A_{TQ_0},     
              \thin A_{UQ_1}, \thin A_{LQ_1}, \thin A_{TQ_1}, 
              \thin A_{UQ_2}, \thin A_{LQ_2}, \thin A_{TQ_2},
\end{eqnarray}
where $U$ denotes the unpolarized case. For example, 
the asymmetry $A_{UQ_0}$ indicates that the proton is unpolarized
and the quadrupole $Q_0$ spin combination is taken for the deuteron.
The precise definitions of these asymmetries should be found in
Ref. \cite{hk1}. The new asymmetries are those with the subscript
$Q_0$, $Q_1$, or $Q_2$.

\section{Parton-model analysis}

The dependence of the structure functions
on the proton and deuteron polarizations 
is revealed by the formalism of the previous section.
However, it is not obvious how these structure functions are
related to the parton distributions.
In particular, the meaning of the new quadrupole structure
functions is not obvious at all. 
In order to clarify it, the hadron tensor is analyzed in
a naive parton model. Because the polarized pd Drell-Yan had
not been discussed before Ref. \cite{hk2}, we should content
ourselves at this stage with the naive analysis:
$O(1/Q)$ contributions are neglected in the course of calculations.

The hadron tensor due to the annihilation process,
$q$(in A)+$\bar q$(in B)$\rightarrow \ell^+ + \ell^-$,
is given in the parton model as
\begin{eqnarray}
& & \! \! \! \! \! \! \! \! \! \! \! \! 
W^{\mu \nu} = \frac{1}{3} \sum_{a, b} \delta_{b \bar{a}} \, e_a^2 
            \int d^4 k_a \, d^4 k_b \, \delta^4 (k_a + k_b - Q)
\nonumber \\
& & \! \! \! \! \! \! \! \! \! \! \! \! 
\times
            Tr [\Phi_{a/A} (P_A S_A; k_a) \gamma^\mu
           \bar{\Phi}_{b/B} (P_B S_B; k_b) \gamma^\nu]
\ ,
\end{eqnarray}
where $k_a$ and $k_b=k_{\bar a}$ are the quark and antiquark momenta,
the color average is taken by the factor $1/3=3\cdot (1/3)^2$,
$e_a$ is the charge of a quark with the flavor $a$, and 
$\Phi$ is a correlation function. Of course, the opposite process
$\bar q$(in A)+$q$(in B)$\rightarrow \ell^+ + \ell^-$ should be
taken into account in order to compare with the experimental cross
section. The correlation function $\Phi (P S; k)$ is a matrix with
sixteen components, so that it can be expanded in terms of
the sixteen $4\times 4$ matrices:
${\bf 1},\, \gamma_5,\, \gamma^\mu,\, \gamma^\mu \gamma_5,\, 
\sigma^{\mu \nu} \gamma_5$ together with the possible Lorentz
vectors and pseudovector: $P^\mu$, $k^\mu$, and $S^\mu$. 
Of course, the expansion terms should satisfy the conditions
of Hermiticity, parity conservation, and time-reversal invariance.
However, the most important point is that the second rank tensors
exist in the deuteron although the spin dependent terms
are allowed up to the linear spin ones in the proton.
It is shown in Ref. \cite{hk2} that the additional terms
give rise to the tensor distribution $b_1$.
We anticipated to have $b_1$; however, we also find a new
one which is related to the intermediate polarization.
It is the first time that we encounter such a distribution,
so that it is simply named a $c_1$ distribution.

Even in the naive analysis, there are still 19 structure functions. 
In order to find the most essential ones,
the cross section is integrated over $\vec Q_T$.
Then, only four finite structure functions exist. Noting that
there are three functions in the pp Drell-Yan, we find a new
structure function which is specific to the deuteron.
Furthermore, we find that it is expressed by the combinations
of unpolarized distributions in the proton with
the tensor polarized distributions in the deuteron.
This structure function can be investigated by
the unpolarized-quadrupole $Q_0$ asymmetry:
\begin{equation}
\! \! \! \! 
A_{UQ_0}  \! =  \! \frac{\sum_a e_a^2 \, 
                  \left[ \, f_1(x_A) \, \bar b_1(x_B)
                          + \bar f_1(x_A) \, b_1(x_B) \, \right] }
                {\sum_a e_a^2 \, 
                  \left[ \, f_1(x_A) \, \bar f_1(x_B)
                          + \bar f_1(x_A) \, f_1(x_B) \, \right] }
. \ 
\label{eqn:auq0}
\end{equation}
The advantage of using the hadron reaction is that the
tensor-polarized antiquark distributions could be obtained
rather easily. In the electron scattering, the antiquark
distributions cannot be determined precisely.
For example, the violation of the Gottfried sum rule
suggested $\bar u \ne \bar d$ \cite{skpr};
however, the precise $x$ dependence
of $\bar u/\bar d$ is determined only recently by the E866
Drell-Yan experiments.
If the large $x_F$ region is considered in Eq. (\ref{eqn:auq0}),
it becomes
\begin{equation}
A_{UQ_0} \textrm{(large $x_F$)} 
      \approx \frac{\sum_a e_a^2 \, f_1(x_A) \, \bar b_1(x_B)}
                   {\sum_a e_a^2 \, f_1(x_A) \, \bar f_1 (x_B)}
\ .
\end{equation}
It indicates that the antiquark tensor distributions $\bar b_1$
can be determined if the unpolarized distributions are well
known in the proton and deuteron.

Another advantage of studying the polarized pd Drell-Yan is
that it becomes possible to extract the flavor asymmetry
$\Delta_T \bar u/\Delta_T \bar d$ in the transversity
distributions by comparing the pp and pd cross sections
\cite{hk1,hk2,km}. 

It is rather difficult to attain the longitudinal polarization
for the deuteron in the collider experiment
(e.g. a possible next-generation RHIC project) because of its
small magnetic moment. However, we could combine the transversely
polarized cross sections with the unpolarized one for investigating
the tensor structure. If a fixed target can be used for the deuteron,
there is no such difficulty. In fact, there are possibilities to
study the polarized pd Drell-Yan at FNAL and also in the HERA-N
project. However, significant theoretical and experimental efforts
are necessary for proposing such an experiment. At this stage,
our numerical analysis is in progress \cite{km} in order to
find the experimental possibilities.

\section{Conclusion}

We discussed first what kinds of structure functions could be studied
in the polarized proton-deuteron Drell-Yan processes. 
Because of the new spin structure for the deuteron,
there exist 108 structure functions. Among them, there are 22
finite ones if the cross section is integrated over $\vec Q_T$.
The new structure functions are associated with the tensor structure
of the deuteron.
The parton-model analysis indicated that there are only four structure
functions in the $\vec Q_T$-integrated case. They are unpolarized,
longitudinally-polarized, transversity, tensor-polarized structure
functions. The last one does not exist in the proton-proton reactions.
It could be measured in the quadrupole $Q_0$ polarization asymmetry
with the unpolarized proton. We expect that the tensor structure will
become one of the exciting topics in high-energy spin physics
in the near future.

\end{document}